\documentclass[11pt]{article}
\usepackage{acl08}
\usepackage{times}
\usepackage{graphicx}
\usepackage{latexsym}
\title{Modeling the Structure and Dynamics of the\\ Consonant Inventories: A Complex Network Approach}

\author{Animesh Mukherjee$^{1}$, Monojit Choudhury$^{2}$, Anupam Basu$^{1}$, Niloy Ganguly$^{1}$\\ $^{1}$Department of 
Computer Science and Engineering,\\ Indian Institute of Technology, Kharagpur, India -- 721302\\ $^{2}$Microsoft 
Research India, Bangalore, India -- 560080\\ {\tt \{animeshm,anupam,niloy\}@cse.iitkgp.ernet.in}, \\{\tt 
monojitc@microsoft.com}}

\date{}

\begin{document}
\maketitle
\begin{abstract}
We study the self-organization of the consonant inventories through a complex network approach. We observe that the 
distribution of occurrence as well as co-occurrence of the consonants across languages follow a power-law behavior. 
The co-occurrence network of consonants exhibits a high clustering coefficient. We propose four novel synthesis models 
for these networks (each of which is a refinement of the earlier) so as to  successively match with higher accuracy 
(a) the above mentioned topological properties as well as (b) the linguistic property of {\em feature economy} 
exhibited by the consonant inventories. We conclude by arguing that a possible interpretation of this mechanism of 
network growth is the process of child language acquisition. Such models essentially increase our understanding of the 
structure of languages that is influenced by their evolutionary dynamics and this, in turn, can be extremely useful 
for building future NLP applications.
\end{abstract}

\section{Introduction}

A large number of regular patterns are observed across the sound inventories of human languages. These regularities 
are arguably a consequence of the self-organization that is instrumental in the emergence of these 
inventories~\cite{Boer:00}. Many attempts have been made by functional phonologists for explaining this 
self-organizing behavior through certain general principles such as {\em maximal perceptual 
contrast}~\cite{Lindblom:72}, {\em ease of articulation}~\cite{Lindblom:88,Boer:00}, and {\em ease of 
learnability}~\cite{Boer:00}. In fact, there are a lot of studies that attempt to explain the emergence of the vowel 
inventories through the application of one or more of the above principles~\cite{Lindblom:72,Boer:00}. Some studies 
have also been carried out in the area of linguistics that seek to reason the observed patterns in the consonant 
inventories~\cite{Trub:39,Lindblom:88,Boersma:98,Clements:04}. Nevertheless, most of these works are confined to 
certain individual principles rather than formulating a general theory describing the emergence of these regular 
patterns across the consonant inventories. 

The self-organization of the consonant inventories emerges due to an interaction of different forces acting upon them. 
In order to identify the nature of these interactions one has to understand the growth dynamics of these inventories. 
The theories of {\em complex networks} provide a number of growth models that have proved to be extremely successful 
in explaining the evolutionary dynamics of various social~\cite{Newman:01,Ramasco:04}, biological~\cite{Jeong:00} and 
other natural systems. The basic framework for the current study develops around two such complex networks namely, the 
{\bf P}honeme-{\bf La}nguage {\bf Net}work or {\bf PlaNet}~\cite{acl:06} and its one-mode projection, the {\bf 
Pho}neme-Phoneme {\bf Net}work or {\bf PhoNet}~\cite{Mukherjee:06}. We begin by analyzing some of the structural 
properties (Sec.~\ref{analysis})  of the networks and observe that the consonant nodes in both PlaNet and PhoNet 
follow a power-law-like degree distribution. Moreover, PhoNet is characterized by a high clustering coefficient, a 
property that has been found to be prevalent in many other social networks~\cite{Newman:01,Ramasco:04}.

We propose four synthesis models for PlaNet (Sec.~\ref{synthesis}), each of which employ a variant of a {\em 
preferential attachment}~\cite{Albert:99} based growth kernel\footnote{The word kernel here refers to the function or 
mathematical formula that drives the growth of the network.}. While the first two models are independent of the 
characteristic properties of the (consonant) nodes, the following two use them. These models are successively refined 
not only to reproduce the topological properties of PlaNet and PhoNet, but also to match the linguistic property of 
{\em feature economy}~\cite{Boersma:98,Clements:04} that is observed across the consonant inventories. The underlying 
growth rules for each of these individual models helps us to interpret the cause of the emergence of at least one (or 
more) of the aforementioned properties. We conclude (Sec.~\ref{conclusion}) by providing a possible interpretation of 
the proposed mathematical model that we finally develop in terms of child language acquisition. 

There are three major contributions of this work. Firstly, it provides a fascinating account of the structure and the 
evolution of the human speech sound systems. Furthermore, the introduction of the node property based synthesis model 
is a significant contribution to the field of complex networks. On a broader perspective, this work shows how 
statistical mechanics can be applied in understanding the structure of a linguistic system, which in turn can be 
extremely useful in developing future NLP applications.

\section{Properties of the Consonant Inventories}\label{analysis}

In this section, we briefly recapitulate the definitions of PlaNet and PhoNet, the data source, construction procedure 
for the networks and some of their important structural properties. We also revisit the concept of feature economy and 
the method used for its quantification.

\subsection{Structural Properties of the Consonant Networks}

PlaNet is a bipartite graph $G$~=~$\langle$~$V_L$, $V_C$, $E_{pl}$~$\rangle$ consisting of two sets of nodes namely, 
$V_L$ (labeled by the languages) and $V_C$ (labeled by the consonants); $E_{pl}$ is the set of edges running between 
$V_L$ and $V_C$. There is an edge $e \in E_{pl}$ from a node $v_l \in$ $V_L$ to a node $v_c \in$ $V_C$ iff the 
consonant $c$ is present in the inventory of language $l$. 

\begin{figure}
\begin{center}
\includegraphics[width=3.2in]{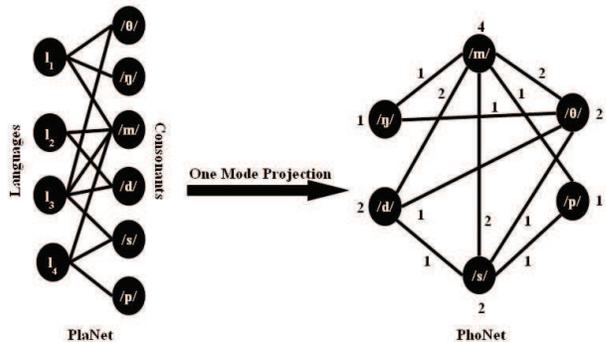}
\caption{Illustration of the nodes and edges of PlaNet and PhoNet.}
\label{pl_ph}
\end{center}
\end{figure}

PhoNet is the one-mode projection of PlaNet onto the consonant nodes i.e., a network of consonants in which two nodes 
are linked by an edge with weight as many times as they co-occur across languages. Hence, it can be represented by a 
graph $G$~=~$\langle$~$V_C$, $E_{ph}$~$\rangle$, where $V_C$ is the set of consonant nodes and $E_{ph}$ is the set of 
edges connecting these nodes in $G$. There is an edge $e \in E_{ph}$ if the two nodes (read consonants) that are 
connected by $e$ co-occur in at least one language and the number of languages they co-occur in defines the weight of 
the edge $e$. Figure~\ref{pl_ph} shows the nodes and the edges of PlaNet and PhoNet.

{\bf Data Source and Network Construction:} Like many other earlier 
studies~\cite{Lindblom:72,Lindblom:88,Boer:00,Hinskens:03}, we use the UCLA Phonological Segment Inventory Database 
(UPSID)~\cite{Maddieson:84} as the source of our data. There are 317 languages in the database with a total of 541 
consonants found across them. Each consonant is characterized by a set of phonological features~\cite{Trub:30}, which 
distinguishes it from others. UPSID uses articulatory features to describe the consonants, which can be broadly 
categorized into three different types namely the {\em manner of articulation}, the {\em place of articulation} and 
{\em phonation}. Manner of articulation specifies how the flow of air takes place in the vocal tract during 
articulation of a consonant, whereas place of articulation specifies the active speech organ and also the place where 
it acts. Phonation describes the vibration of the vocal cords during the articulation of a consonant. Apart from these 
three major classes there are also some secondary articulatory features found in certain languages. There are around 
52 features listed in UPSID; the important ones are noted in Table~\ref{tab0}. Note that in UPSID the features are 
assumed to be binary-valued and therefore, each consonant can be represented by a binary vector. 

We have used UPSID in order to construct PlaNet and PhoNet. Consequently, $\mid$$V_L$$\mid$ = 317 (in PlaNet) and 
$\mid$$V_C$$\mid$ = 541. The number of edges in PlaNet and PhoNet are 7022 and 30412 respectively.     

\begin{table}\centering
\resizebox{!}{1.0in}{
\begin{tabular}{lll}
\hline
\vbox to2.04ex{\vspace{1pt}\vfil\hbox to22.80ex{\hfil Manner of Articulation\hfil}} & 
\vbox to2.04ex{\vspace{1pt}\vfil\hbox to22.40ex{\hfil Place of Articulation\hfil}} & 
\vbox to2.04ex{\vspace{1pt}\vfil\hbox to21.00ex{\hfil Phonation\hfil}} \\

\hline
\hline
\vbox to1.88ex{\vspace{1pt}\vfil\hbox to22.80ex{\hfil tap\hfil}} & 
\vbox to1.88ex{\vspace{1pt}\vfil\hbox to22.40ex{\hfil velar\hfil}} & 
\vbox to1.88ex{\vspace{1pt}\vfil\hbox to21.00ex{\hfil voiced\hfil}} \\

\vbox to1.88ex{\vspace{1pt}\vfil\hbox to22.80ex{\hfil flap\hfil}} & 
\vbox to1.88ex{\vspace{1pt}\vfil\hbox to22.40ex{\hfil uvular\hfil}} & 
\vbox to1.88ex{\vspace{1pt}\vfil\hbox to21.00ex{\hfil voiceless\hfil}} \\

\vbox to1.88ex{\vspace{1pt}\vfil\hbox to22.80ex{\hfil trill\hfil}} & 
\vbox to1.88ex{\vspace{1pt}\vfil\hbox to22.40ex{\hfil dental\hfil}} & 
\vbox to1.88ex{\vspace{1pt}\vfil\hbox to21.00ex{\hfil \hfil}} \\

\vbox to1.88ex{\vspace{1pt}\vfil\hbox to22.80ex{\hfil click\hfil}} & 
\vbox to1.88ex{\vspace{1pt}\vfil\hbox to22.40ex{\hfil palatal\hfil}} & 
\vbox to1.88ex{\vspace{1pt}\vfil\hbox to21.00ex{\hfil \hfil}} \\

\vbox to1.88ex{\vspace{1pt}\vfil\hbox to22.80ex{\hfil nasal\hfil}} & 
\vbox to1.88ex{\vspace{1pt}\vfil\hbox to22.40ex{\hfil glottal\hfil}} & 
\vbox to1.88ex{\vspace{1pt}\vfil\hbox to21.00ex{\hfil \hfil}} \\

\vbox to1.88ex{\vspace{1pt}\vfil\hbox to22.80ex{\hfil plosive\hfil}} & 
\vbox to1.88ex{\vspace{1pt}\vfil\hbox to22.40ex{\hfil bilabial\hfil}} & 
\vbox to1.88ex{\vspace{1pt}\vfil\hbox to21.00ex{\hfil \hfil}} \\

\vbox to1.88ex{\vspace{1pt}\vfil\hbox to22.80ex{\hfil r-sound\hfil}} & 
\vbox to1.88ex{\vspace{1pt}\vfil\hbox to22.40ex{\hfil alveolar\hfil}} & 
\vbox to1.88ex{\vspace{1pt}\vfil\hbox to21.00ex{\hfil \hfil}} \\

\vbox to1.88ex{\vspace{1pt}\vfil\hbox to22.80ex{\hfil fricative\hfil}} & 
\vbox to1.88ex{\vspace{1pt}\vfil\hbox to22.40ex{\hfil retroflex\hfil}} & 
\vbox to1.88ex{\vspace{1pt}\vfil\hbox to21.00ex{\hfil \hfil}} \\

\vbox to1.88ex{\vspace{1pt}\vfil\hbox to22.80ex{\hfil affricate\hfil}} & 
\vbox to1.88ex{\vspace{1pt}\vfil\hbox to22.40ex{\hfil pharyngeal\hfil}} & 
\vbox to1.88ex{\vspace{1pt}\vfil\hbox to21.00ex{\hfil \hfil}} \\

\vbox to1.88ex{\vspace{1pt}\vfil\hbox to22.80ex{\hfil implosive\hfil}} & 
\vbox to1.88ex{\vspace{1pt}\vfil\hbox to22.40ex{\hfil labial-velar\hfil}} & 
\vbox to1.88ex{\vspace{1pt}\vfil\hbox to21.00ex{\hfil \hfil}} \\

\vbox to1.88ex{\vspace{1pt}\vfil\hbox to22.80ex{\hfil approximant\hfil}} & 
\vbox to1.88ex{\vspace{1pt}\vfil\hbox to22.40ex{\hfil labio-dental\hfil}} & 
\vbox to1.88ex{\vspace{1pt}\vfil\hbox to21.00ex{\hfil \hfil}} \\

\vbox to1.88ex{\vspace{1pt}\vfil\hbox to22.80ex{\hfil ejective stop\hfil}} & 
\vbox to1.88ex{\vspace{1pt}\vfil\hbox to22.40ex{\hfil labial-palatal\hfil}} & 
\vbox to1.88ex{\vspace{1pt}\vfil\hbox to21.00ex{\hfil \hfil}} \\

\vbox to1.88ex{\vspace{1pt}\vfil\hbox to22.80ex{\hfil affricated click\hfil}} & 
\vbox to1.88ex{\vspace{1pt}\vfil\hbox to22.40ex{\hfil dental-palatal\hfil}} & 
\vbox to1.88ex{\vspace{1pt}\vfil\hbox to21.00ex{\hfil \hfil}} \\

\vbox to1.88ex{\vspace{1pt}\vfil\hbox to22.80ex{\hfil ejective affricate\hfil}} & 
\vbox to1.88ex{\vspace{1pt}\vfil\hbox to22.40ex{\hfil dental-alveolar\hfil}} & 
\vbox to1.88ex{\vspace{1pt}\vfil\hbox to21.00ex{\hfil \hfil}} \\

\vbox to1.88ex{\vspace{1pt}\vfil\hbox to22.80ex{\hfil ejective fricative\hfil}} & 
\vbox to1.88ex{\vspace{1pt}\vfil\hbox to22.40ex{\hfil palato-alveolar\hfil}} & 
\vbox to1.88ex{\vspace{1pt}\vfil\hbox to21.00ex{\hfil \hfil}} \\

\vbox to1.88ex{\vspace{1pt}\vfil\hbox to22.80ex{\hfil lateral approximant\hfil}} & 
\vbox to1.88ex{\vspace{1pt}\vfil\hbox to22.40ex{\hfil \hfil}} & 
\vbox to1.88ex{\vspace{1pt}\vfil\hbox to21.00ex{\hfil \hfil}} \\

\cline{1-3}
\end{tabular}}
\caption{The table shows some of the important features listed in UPSID. Over 99\% of the UPSID languages have 
bilabial, dental-alveolar and velar plosives. Furthermore, voiceless plosives outnumber the voiced ones (92\% vs. 
67\%). 93\% of the languages have at least one fricative, 97\% have at least one nasal and 96\% have at least one 
liquid. Approximants occur in fewer than 95\% of the languages.}
\label{tab0}
\end{table}

\begin{figure*}[!t]
\begin{center}
\includegraphics[width=6in]{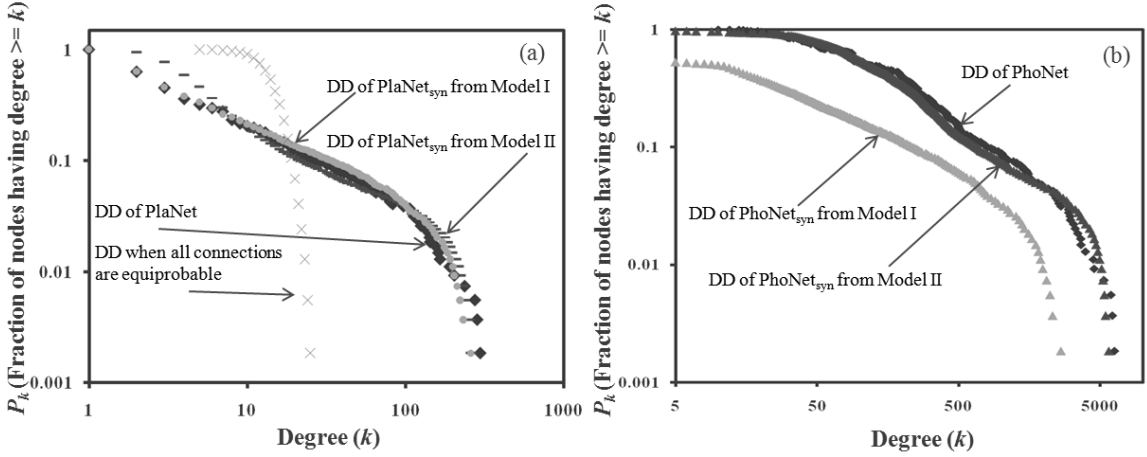}
\caption{Degree distribution (DD) of PlaNet along with that of PlaNet$_{syn}$ obtained from Model I and II 
respectively; (b) DD of PhoNet along with that of PhoNet$_{syn}$ obtained from Model I and II respectively. Both the 
plots are in log-log scale.}
\label{dd}
\end{center}
\end{figure*}

{\bf Degree Distributions of PlaNet and PhoNet:} The degree distribution is the fraction of nodes, denoted by $P_k$, 
which have a degree\footnote{For a weighted graph like PhoNet, the degree of a node $i$ is the sum of weights on the 
edges that are incident on $i$.} greater than or equal to $k$~\cite{Newman:03}. The degree distribution of the 
consonant nodes in PlaNet and PhoNet are shown in Figure~\ref{dd} in the log-log scale. Both the plots show a 
power-law behavior ($P_k \propto k^{-\alpha}$) with exponential cut-offs  towards the ends. The value of $\alpha$ is 
0.71 for PlaNet and 0.89 for PhoNet.
 
{\bf Clustering Coefficient of PhoNet:} The clustering coefficient for a node $i$ is the proportion of links between 
the nodes that are the neighbors of $i$ divided by the number of links that could possibly exist between 
them~\cite{Newman:03}. Since PhoNet is a weighted graph the above definition is suitably modified by the one presented 
in~\cite{Barrat:04}. According to this definition, the clustering coefficient for a node $i$ is, 
\begin{equation}\label{cc} c_i = \frac{1}{\left(\sum_{\forall j}w_{ij}\right)(k_i-1)}{\sum_{\forall j, 
l}\frac{(w_{ij}+w_{il})}{2}a_{ij}a_{il}a_{jl}}\end{equation} where $j$ and $l$ are neighbors of $i$; $k_i$ represents 
the plain degree of the node $i$; $w_{ij}$, $w_{jl}$ and $w_{il}$ denote the weights of the edges connecting nodes $i$ 
and $j$, $j$ and $l$, and $i$ and $l$ respectively; $a_{ij}$, $a_{il}$, $a_{jl}$ are boolean variables, which are true 
iff there is an edge between the nodes $i$ and $j$, $i$ and $l$, and $j$ and $l$ respectively. The clustering 
coefficient of the network ($c_{av}$) is equal to the average clustering coefficient of the nodes. The value of 
$c_{av}$ for PhoNet is 0.89, which is significantly higher than that of a random graph with the same number of nodes 
and edges ($0.08$).

\subsection{Linguistic Properties: Feature Economy and its Quantification}

The principle of feature economy states that languages tend to use a small number of {\em distinctive} features and 
maximize their combinatorial possibilities to generate a large number of consonants~\cite{Boersma:98,Clements:04}. 
Stated differently, a given consonant will have a higher than expected chance of occurrence in inventories in which 
all of its features have already distinctively occurred in the other consonants.  This principle immediately implies 
that the consonants chosen by a language should share a considerable number of features among them. The quantification 
process, which is a refinement of the idea presented in~\cite{acl:07}, is as follows.

{\bf Feature Entropy:} For an inventory of size $N$, let there be $p_f$ consonants for which a particular feature $f$ 
(recall that we assume $f$ to be binary-valued) is present and $q_f$ other consonants for which the same is absent. 
Therefore, the probability that a consonant (chosen uniformly at random from this inventory) contains the feature $f$ 
is  $\frac {p_f}{N}$ and the probability that it does not contain the feature is $\frac {q_f}{N}$ 
(=1--$\frac{p_f}{N}$). One can think of $f$ as an independent random variable, which can take values 1 and 0, and 
$\frac{p_f}{N}$ and $\frac{q_f}{N}$ define the probability distribution of $f$. Therefore, for any given inventory, we 
can define the  binary entropy $H_f$~\cite{Shan:49} for the feature $f$ as 
\begin{equation}
H_f = - \frac{p_f}{N}\log_2 \frac{p_f}{N} - \frac{q_f}{N}\log_2 \frac{q_f}{N}
\end{equation} 
If $F$ is the set of all features present in the consonants forming the inventory, then {\em feature entropy} $F_E$ is 
the sum of the binary entropies with respect to all the features, that is
\begin{equation}\label{eqfe}
F_E = \sum_{f \in F} H_f = \sum_{f \in F}(- \frac {p_f}{N}\log_2{\frac{p_f}{N}} - \frac{q_f}{N}\log_2{\frac {q_f}{N}})  
\end{equation} 


Since we have assumed that $f$ is an independent random variable, $F_E$ is the joint entropy of the system. In other 
words, $F_E$ provides an estimate of the number of discriminative features present in the consonants of an inventory 
that a speaker (e.g., parent) has to communicate to a learner (e.g., child) during language transmission. The lower 
the value of $F_E$ the higher is the feature economy. The curve marked as (R) in Figure~\ref{fe} shows the average 
feature entropy of the consonant inventories of a particular size\footnote{Let there be $n$ inventories of a 
particular size $k$. The average feature entropy of the inventories of size $k$ is 
$\frac{1}{n}{\sum_{i=1}^n{F_{E_i}}}$, where $F_{E_i}$ signifies the feature entropy of the $i^{th}$ inventory of size 
$k$.} (y-axis) versus the inventory size (x-axis).

\section{Synthesis Models}\label{synthesis}

In this section, we describe four synthesis models that incrementally attempt to explain the emergence of the 
structural properties of PlaNet and PhoNet as well as the feature entropy exhibited by the consonant inventories. In 
all these models, we assume that the distribution of the consonant inventory size, i.e., the degrees of the language 
nodes in PlaNet, are known {\em a priori}. 

\subsection{Model I: Preferential Attachment Kernel}

This model employs a modified version of the kernel described in~\cite{acl:06}, which is the only work in literature 
that attempts to explain the emergence of the consonant inventories in the framework of complex networks.

Let us assume that a language node $L_i \in$ $V_L$ has a degree $k_i$. The consonant nodes in $V_C$ are assumed to be 
unlabeled, i.e, they are not marked by the distinctive features that characterize them. We first sort the nodes $L_1$ 
through $L_{317}$ in the ascending order of their degrees. At each time step a node $L_j$, chosen in order, 
preferentially attaches itself with $k_j$ {\em distinct} nodes (call each such node $C_i$) of the set $V_C$. The 
probability $Pr(C_i)$ with which the node $L_j$ attaches itself to the node $C_i$ is given by, 
\begin{equation}\label{pref}
Pr(C_i) = \frac{{d_{i}}^{\alpha} + \epsilon}{\sum_{{i^{'}} \in V^{'}_{C}} ({d_{i^{'}}}^{\alpha} + \epsilon)}
\end{equation} where, $d_i$ is the current degree of the node $C_i$, $V^{'}_{C}$ is the set of nodes in $V_C$ that are 
not already connected to $L_j$, $\epsilon$ is the smoothing parameter that facilitates random attachments and $\alpha$ 
indicates whether the attachment kernel is sub-linear ($\alpha < 1$), linear ($\alpha = 1$) or super-linear ($\alpha > 
1$). Note that the modification from the earlier kernel~\cite{acl:06} is brought about by the introduction of 
$\alpha$. The above process is repeated until all the language nodes $L_j \in V_L$ get connected to $k_j$ consonant 
nodes (refer to Figure. 6 of~\cite{acl:06} for an illustration of the steps of the synthesis process). Thus, we have 
the synthesized version of PlaNet, which we shall call PlaNet$_{syn}$ henceforth.

{\bf The Simulation Results:} We simulate the above model to obtain PlaNet$_{syn}$ for 100 different runs and average 
the results over all of them. We find that the degree distributions that emerge fit the empirical data well for 
$\alpha \in$ [1.4,1.5] and $\epsilon \in$ [0.4,0.6], the best being at $\alpha$~=~1.44 and $\epsilon$~=~0.5 (shown in 
Figure~\ref{dd}). In fact, the mean error\footnote{Mean error is defined as the average difference between the 
ordinate pairs (say $y$ and $y^{'}$) where the abscissas are equal. In other words, if there are $N$ such ordinate 
pairs then the mean error can be expressed as $\frac{\sum\mid y-y^{'}\mid}{N}$.} between the real and the synthesized 
distributions for the best choice of parameters is as small as 0.01. Note that this error in case of the model 
presented in~\cite{acl:06} was 0.03. Furthermore, as we shall see shortly, a super-linear kernel can explain various 
other topological properties more accurately than a linear kernel. 

In absence of preferential attachment i.e., when all the connections to the consonant nodes are equiprobable, the mean 
error rises to 0.35. 

A possible reason behind the success of this model is the fact that language is a constantly changing system and 
preferential attachment plays a significant role in this change. For instance, during the change those consonants that 
belong to languages that are more prevalent among the speakers of a generation have higher chances of being 
transmitted to the speakers of the subsequent generations~\cite{Blevins:04}. This heterogeneity in the choice of the 
consonants manifests itself as preferential attachment. We conjecture that the value of $\alpha$ is a function of the 
societal structure and the cognitive capabilities of human beings. The exact nature of this function is currently not 
known and a topic for future research. The parameter $\epsilon$ in this case may be thought of as modeling the 
randomness of the system.

Nevertheless, the degree distribution of PhoNet$_{syn}$, which is the one-mode projection of PlaNet$_{syn}$, does not 
match the real data well (see Figure~\ref{dd}). The mean error between the two distributions is 0.45. Furthermore, the 
clustering coefficient of PhoNet$_{syn}$ is 0.55 and differs largely from that of PhoNet. The primary reason for this 
deviation in the results is that PhoNet exhibits strong patterns of co-occurrences~\cite{Mukherjee:06} and this fact 
is not taken into account by Model I. In order to circumvent the above problem, we introduce the concept of {\em 
triad} (i.e., fully connected triplet) formation and thereby refine the model in the following section.   


\subsection{Model II: Kernel based on Triad Formation}

The triad model~\cite{Alava:06} builds up on the concept of {\em neighborhood} formation. Two consonant nodes $C_1$ 
and $C_2$ become neighbors if a language node at any step of the synthesis process attaches itself to both $C_1$ and 
$C_2$. Let the probability of triad formation be denoted by $p_t$. At each time step a language node $L_j$ (chosen 
from the set of language nodes sorted in ascending order of their degrees) makes the first connection preferentially 
to a consonant node $C_i$ $\in$ $V_C$ to which $L_j$ is not already connected following the distribution $Pr(C_i)$. 
For the rest of the ($k_j$-1) connections $L_j$ attaches itself preferentially to only the neighbors of $C_i$ to which 
$L_j$ is not yet connected with a probability $p_t$. Consequently, $L_j$ connects itself preferentially to the 
non-neighbors of $C_i$ to which $L_j$ is not yet connected with a probability ($1-p_t$). The neighbor set of $C_i$ 
gets updated accordingly. Note that every time the node $C_i$ and its neighbors are chosen they together impose a 
clique on the one-mode projection. This phenomenon leads to the formation of a large number of triangles in the 
one-mode projection thereby increasing the clustering coefficient of the resultant network. 

{\bf The Simulation Results:} We carry out 100 different simulation runs of the above model for a particular set of 
parameter values to obtain PlaNet$_{syn}$ and average the results over all of them. We explore several parameter 
settings in the range as follows: $\alpha \in$ [1,1.5] (in steps of 0.1), $\epsilon \in$ [0.2,0.4] (in steps of 0.1) 
and $p_t \in$ [0.70,0.95] (in steps of 0.05). We also observe that if we traverse any further along one or more of the 
dimensions of the parameter space then the results get worse. The best result emerges for $\alpha$~=~1.3, 
$\epsilon$~=~0.3 and $p_t$~=~0.8. 

Figure~\ref{dd} shows the degree distribution of the consonant nodes of PlaNet$_{syn}$ and PlaNet. The mean error 
between the two distributions is 0.04 approximately and is therefore worse than the result obtained from Model I. 
Nevertheless, the average clustering coefficient of PhoNet$_{syn}$ in this case is 0.85, which is within 4.5\% of that 
of PhoNet. Moreover, in this process the mean error between the degree distribution of PhoNet$_{syn}$ and PhoNet (as 
illustrated in Figure~\ref{dd}) has got reduced drastically from 0.45 to 0.03.

One can again find a possible association of this model with the phenomena of language change. If a group of 
consonants largely co-occur in the languages of a generation of speakers then it is very likely that all of them get 
transmitted together in the subsequent generations~\cite{Blevins:04}. The triad formation probability ensures that if 
a pair of consonant nodes become neighbors of each other in a particular step of the synthesis process then the choice 
of such a pair should be highly preferred in the subsequent steps of the process. This is coherent with the 
aforementioned phenomenon of transmission of consonants in groups over linguistic generations. Since the value of 
$p_t$ that we obtain is quite high, it may be argued that such transmissions are largely prevalent in nature. 


Although Model II reproduces the structural properties of PlaNet and PhoNet quite accurately, as we shall see shortly, 
it fails to generate inventories that closely match the real ones in terms of feature entropy. However, at this point, 
recall that Model II assumes that the consonant nodes are unlabeled; therefore, the inventories that are produced as a 
result of the synthesis are composed of consonants, which unlike the real inventories, are not marked by their 
distinctive features. In order to label them we perform the following,\\
{\bf The Labeling Scheme:}\\
1. Sort the consonants of UPSID in the decreasing order of their frequency of occurrence and call this list of 
consonants $ListC[1 \cdot \cdot \cdot 541]$,\\
2. Sort the $V_C$ nodes of PlaNet$_{syn}$ in decreasing order of their degree and call this list of nodes $ListN[1 
\cdot \cdot \cdot 541]$,\\
3. $\forall_{1 \le i \le 541}$ $ListN[i]\longleftarrow ListC[i]$  
     
\begin{figure}
\begin{center}
\includegraphics[width=3in]{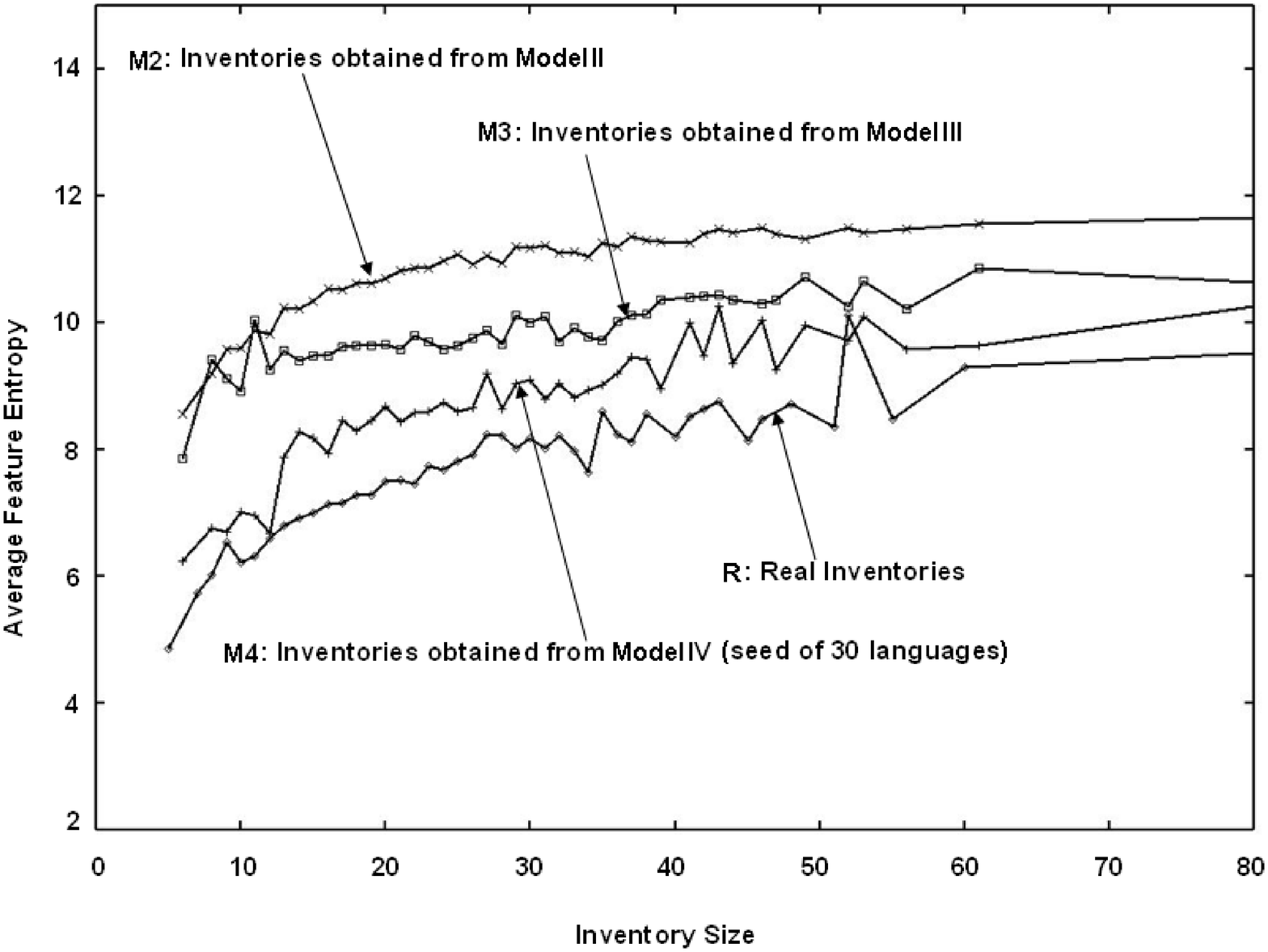}
\caption{Average feature entropy of the inventories of a particular size (y-axis) versus the inventory size (x-axis).}
\label{fe}
\end{center}
\end{figure}

The Figure~\ref{fe} indicates that the curve for the real inventories (R) and those obtained from Model II (M2) are 
significantly different from each other. This difference arises due to the fact that in Model II, the choice of a 
consonant from the set of neighbors is solely degree-dependent, where the relationships between the features are not 
taken into consideration. Therefore, in order to eliminate this problem, we introduce the model using the 
feature-based kernel in the next section.  

\subsection{Model III: Feature-based Kernel}

In this model, we assume that each of the consonant nodes are labeled, that is each of them are marked by a set of 
distinctive features. The attachment kernel in this case has two components one of which is preferential while the 
other favors the choice of those consonants that are at a low feature distance (the number of feature positions they 
differ at) from the already chosen ones. Let us denote the feature distance between two consonants $C_i$ and $C^{'}_i$ 
by $D(C_i,C^{'}_i)$. We define the {\em affinity}, $A(C_i,C^{'}_i)$, between $C_i$ and $C^{'}_i$ as
\begin{equation}
A(C_i,C^{'}_i) = \frac{1}{D(C_i,C^{'}_i)}
\end{equation} 
Therefore, the lower the feature distance between $C_i$ and $C^{'}_i$ the higher is the affinity between them.
 
At each time step a language node establishes the first connection with a consonant node (say $C_i$) preferentially 
following the distribution $Pr(C_i)$ like the previous models. The rest of the connections to any arbitrary consonant 
node $C^{'}_i$ (not yet connected to the language node) are made following the distribution $(1-w)Pr(C^{'}_i) + 
wPr_{aff}(C_i, C^{'}_i)$, where 
\begin{equation}
Pr_{aff}(C_i, C^{'}_i) = \frac{A(C_i, C^{'}_i)}{\sum_{\forall C^{'}_i}A(C_i, C^{'}_i)}
\end{equation}
and $0 < w <1$.

{\bf Simulation Results:} We perform 100 different simulation runs of the above model for a particular set of 
parameter values to obtain PlaNet$_{syn}$ and average the results over all of them. We explore different parameter 
settings in the range as follows: $\alpha \in$ [1,2] (in steps of 0.1), $\epsilon \in$ [0.1,1] (in steps of 0.1) and 
$w \in$ [0.1,0.5] (in steps of 0.05). The best result in terms of the structural properties of PlaNet and PhoNet 
emerges for $\alpha$~=~1.6, $\epsilon$~=~0.3 and $w$~=~0.2. 


In this case, the mean error between the degree distribution curves for PlaNet$_{syn}$ and PlaNet is 0.05 and that 
between of PhoNet$_{syn}$ and PhoNet is 0.02. Furthermore, the clustering coefficient of PhoNet$_{syn}$ in this case 
is 0.84, which is within 5.6\% of that of PhoNet. The above results show that the structural properties of the 
synthesized networks in this case are quite similar to those obtained through the triad model. Nevertheless, the 
average feature entropy of the inventories produced (see curve M3 in Figure~\ref{fe}) are more close to that of the 
real ones now (for quantitative comparison see Table~\ref{tabf}). 

Therefore, it turns out that the groups of consonants that largely co-occur in the languages of a linguistic 
generation are actually driven by the principle of feature economy (see~\cite{Clements:04,Mukherjee:06} for details).   

However, note that even for Model III the nodes that are chosen for attachment in the initial stages of the synthesis 
process are arbitrary and consequently, the labels of the nodes of PlaNet$_{syn}$ do not have a one-to-one 
correspondence with that of PlaNet, which is the main reason behind the difference in the result between them. In 
order to overcome this problem we can make use of a small set of real inventories to bootstrap the model. 

\subsection{Model IV: Feature-based Kernel and Bootstrapping}

In order to create a bias towards the labeling scheme prevalent in PlaNet, we use 30 (around 10\% of the) real 
languages as a seed (chosen randomly) for Model III; i.e., they are used by the model for bootstrapping. The idea is 
summarized below.\\
1. Select 30 real inventories at random and construct a PlaNet from them. Call this network the initial 
PlaNet$_{syn}$.\\
2. The rest of the language nodes are incrementally added to this initial PlaNet$_{syn}$ using Model III. 

{\bf Simulation Results:} The best fit now emerges at $\alpha = 1.35$, $\epsilon = 0.3$ and $w = 0.15$. The mean error 
between the degree distribution of PlaNet and PlaNet$_{syn}$ is 0.05 and that between PhoNet and PhoNet$_{syn}$ is 
0.02. The clustering coefficient of PhoNet$_{syn}$ is 0.83 in this case (within 6.7\% of that of PhoNet).

The inventories that are produced as a result of the bootstrapping have an average feature entropy closer to the real 
inventories (see curve M4 in Figure~\ref{fe}) than the earlier models. Hence, we find that this improved labeling 
strategy brings about a global betterment in our results unlike in the previous cases. The larger the number of 
languages used for the purpose of bootstrapping the better are the results mainly in terms of the match in the feature 
entropy curves. 

\begin{table}\centering
\resizebox{!}{0.52in}{
\begin{tabular}{lllllllll}
\hline
\vbox to1.88ex{\vspace{1pt}\vfil\hbox to30.80ex{\hfil Results\hfil}} & 
\vbox to1.88ex{\vspace{1pt}\vfil\hbox to7.00ex{\hfil Model I\hfil}} & 
\vbox to1.88ex{\vspace{1pt}\vfil\hbox to7.00ex{\hfil Model II\hfil}} & 
\vbox to1.88ex{\vspace{1pt}\vfil\hbox to7.40ex{\hfil Model III\hfil}} & 
\vbox to1.88ex{\vspace{1pt}\vfil\hbox to7.60ex{\hfil Model IV\hfil}} \\

\hline
\hline
\vbox to1.88ex{\vspace{1pt}\vfil\hbox to30.80ex{\hfil ME: DD of PlaNet \& PlaNet$_{syn}$\hfil}} & 
\vbox to1.88ex{\vspace{1pt}\vfil\hbox to7.00ex{\hfil 0.01\hfil}} & 
\vbox to1.88ex{\vspace{1pt}\vfil\hbox to7.00ex{\hfil 0.04\hfil}} & 
\vbox to1.88ex{\vspace{1pt}\vfil\hbox to7.40ex{\hfil 0.05\hfil}} & 
\vbox to1.88ex{\vspace{1pt}\vfil\hbox to7.60ex{\hfil 0.05\hfil}} \\

\vbox to1.88ex{\vspace{1pt}\vfil\hbox to30.80ex{\hfil ME: DD of PhoNet \& PhoNet$_{syn}$\hfil}} & 
\vbox to1.88ex{\vspace{1pt}\vfil\hbox to7.00ex{\hfil 0.45\hfil}} & 
\vbox to1.88ex{\vspace{1pt}\vfil\hbox to7.00ex{\hfil 0.03\hfil}} & 
\vbox to1.88ex{\vspace{1pt}\vfil\hbox to7.40ex{\hfil 0.02\hfil}} & 
\vbox to1.88ex{\vspace{1pt}\vfil\hbox to7.60ex{\hfil 0.02\hfil}} \\

\vbox to1.88ex{\vspace{1pt}\vfil\hbox to30.80ex{\hfil \% Err: Clustering Coefficient\hfil}} & 
\vbox to1.88ex{\vspace{1pt}\vfil\hbox to7.00ex{\hfil 38.2\hfil}} & 
\vbox to1.88ex{\vspace{1pt}\vfil\hbox to7.00ex{\hfil 04.5\hfil}} & 
\vbox to1.88ex{\vspace{1pt}\vfil\hbox to7.40ex{\hfil 05.6\hfil}} & 
\vbox to1.88ex{\vspace{1pt}\vfil\hbox to7.60ex{\hfil 06.7\hfil}} \\

\vbox to1.88ex{\vspace{1pt}\vfil\hbox to30.80ex{\hfil ME: Avg. F$_E$ of Real \& Synth. Inv.\hfil}} & 
\vbox to1.88ex{\vspace{1pt}\vfil\hbox to7.00ex{\hfil 3.40\hfil}} & 
\vbox to1.88ex{\vspace{1pt}\vfil\hbox to7.00ex{\hfil 3.00\hfil}} & 
\vbox to1.88ex{\vspace{1pt}\vfil\hbox to7.40ex{\hfil 2.10\hfil}} & 
\vbox to1.88ex{\vspace{1pt}\vfil\hbox to7.60ex{\hfil 0.93\hfil}} \\

\vbox to1.88ex{\vspace{1pt}\vfil\hbox to30.80ex{\hfil $\alpha$\hfil}} & 
\vbox to1.88ex{\vspace{1pt}\vfil\hbox to7.00ex{\hfil 1.44\hfil}} & 
\vbox to1.88ex{\vspace{1pt}\vfil\hbox to7.00ex{\hfil 1.30\hfil}} & 
\vbox to1.88ex{\vspace{1pt}\vfil\hbox to7.40ex{\hfil 1.60\hfil}} & 
\vbox to1.88ex{\vspace{1pt}\vfil\hbox to7.60ex{\hfil 1.35\hfil}} \\

\vbox to1.88ex{\vspace{1pt}\vfil\hbox to30.80ex{\hfil $\epsilon$\hfil}} & 
\vbox to1.88ex{\vspace{1pt}\vfil\hbox to7.00ex{\hfil 0.5\hfil}} & 
\vbox to1.88ex{\vspace{1pt}\vfil\hbox to7.00ex{\hfil 0.3\hfil}} & 
\vbox to1.88ex{\vspace{1pt}\vfil\hbox to7.40ex{\hfil 0.3\hfil}} & 
\vbox to1.88ex{\vspace{1pt}\vfil\hbox to7.60ex{\hfil 0.3\hfil}} \\

\vbox to1.88ex{\vspace{1pt}\vfil\hbox to30.80ex{\hfil $p_t$\hfil}} & 
\vbox to1.88ex{\vspace{1pt}\vfil\hbox to7.00ex{\hfil --\hfil}} & 
\vbox to1.88ex{\vspace{1pt}\vfil\hbox to7.00ex{\hfil 0.8\hfil}} & 
\vbox to1.88ex{\vspace{1pt}\vfil\hbox to7.40ex{\hfil --\hfil}} & 
\vbox to1.88ex{\vspace{1pt}\vfil\hbox to7.60ex{\hfil --\hfil}} \\

\vbox to1.88ex{\vspace{1pt}\vfil\hbox to30.80ex{\hfil $w$\hfil}} & 
\vbox to1.88ex{\vspace{1pt}\vfil\hbox to7.00ex{\hfil --\hfil}} & 
\vbox to1.88ex{\vspace{1pt}\vfil\hbox to7.00ex{\hfil --\hfil}} & 
\vbox to1.88ex{\vspace{1pt}\vfil\hbox to7.40ex{\hfil 0.20\hfil}} & 
\vbox to1.88ex{\vspace{1pt}\vfil\hbox to7.60ex{\hfil 0.15\hfil}} \\
\cline{1-5}
\end{tabular}}
\caption{Important results obtained from each of the models. ME: Mean Error, DD: Degree Distribution.}
\label{tabf}
\end{table} 

\section{Conclusion}\label{conclusion}

We dedicated the preceding sections of this article to analyze and synthesize the consonant inventories of the world's 
languages in the framework of a complex network. Table~\ref{tabf} summarizes the results obtained from the four models 
so that the reader can easily compare them. Some of our important observations are\\
$\bullet$ The distribution of occurrence and co-occurrence of consonants across languages roughly follow a power 
law,\\
$\bullet$ The co-occurrence network of consonants has a large clustering coefficient,\\
$\bullet$ Groups of consonants that largely co-occur across languages are driven by feature economy (which can be 
expressed through feature entropy),\\  
$\bullet$ Each of the above properties emerges due to different reasons, which are successively unfurled by our 
models. 

So far, we have tried to explain the physical significance of our models in terms of the process of language change. 
Language change is a collective phenomenon that functions at the level of a population of speakers~\cite{Steels:00}. 
Nevertheless, it is also possible to explain the significance of the models at the level of an individual, primarily 
in terms of the process of language acquisition, which largely governs the course of language change. In the initial 
years of language development every child passes through a stage called {\em babbling} during which he/she learns to 
produce non-meaningful sequences of consonants and vowels, some of which are not even used in the language to which 
they are exposed~\cite{Jakob:68,Locke:83}. Clear preferences can be observed for learning certain sounds such as 
plosives and nasals, whereas fricatives and liquids are avoided. In fact, this hierarchy of preference during the 
babbling stage follows the cross-linguistic frequency distribution of the consonants. This innate frequency dependent 
preference towards certain phonemes might be because of phonetic reasons (i.e., for articulatory/perceptual benefits). 
In all our models, this innate preference gets captured through the process of preferential attachment. However, at 
the same time, in the context of learning a particular inventory the ease of learning the individual consonants also 
plays an important role. The lower the number of new feature distinctions to be learnt, the higher the ease of 
learning the consonant. Therefore, there are two orthogonal preferences: (a) the occurrence frequency dependent 
preference (that is innate), and (b) the feature-dependent preference (that increases the ease of learning), which are 
instrumental in the acquisition of the inventories. The feature-based kernel is essentially a linear combination of 
these two mutually orthogonal factors.


\begin{thebibliography}{}

\bibitem[\protect\citename{Barab{\'a}si and Albert}1999]{Albert:99}
A.-L. Barab{\'a}si and R. Albert.
\newblock 1999.
\newblock Emergence of scaling in random networks. {\em Science} {\bf 286}, 509-–512.

\bibitem[\protect\citename{Barrat \bgroup et al.\egroup}2004]{Barrat:04}
A. Barrat, M. Barth{\'e}lemy, R. Pastor-Satorras and A. Vespignani.
\newblock 2004.
\newblock The architecture of complex weighted networks. {\em PNAS} {\bf 101}, 3747--3752.

\bibitem[\protect\citename{Blevins}2004]{Blevins:04}
J. Blevins.
\newblock 2004.
\newblock {\em Evolutionary Phonology: The Emergence of Sound Patterns}, Cambridge University Press, Cambridge.

\bibitem[\protect\citename{de Boer}2000]{Boer:00}
B. de Boer.
\newblock 2000.
\newblock Self-organisation in vowel systems. {\em Journal of Phonetics} {\bf 28}(4), 441--465.

\bibitem[\protect\citename{Boersma}1998]{Boersma:98}
P. Boersma.
\newblock 1998.  
\newblock {\em Functional Phonology}, The Hague: Holland Academic Graphics.

\bibitem[\protect\citename{Choudhury \bgroup et al.\egroup}2006]{acl:06}
M. Choudhury, A. Mukherjee, A. Basu and N. Ganguly.
\newblock 2006.
\newblock Analysis and synthesis of the distribution of consonants over languages: A complex network approach. {\em 
Proceedings of COLING-ACL06}, 128--135.

\bibitem[\protect\citename{Clements}2008]{Clements:04}
G. N. Clements.
\newblock 2008.
\newblock The role of features in speech sound inventories. In Eric Raimy \& Charles Cairns, eds.,{\em Contemporary 
Views on Architecture and Representations in Phonological Theory}, Cambridge, MA: MIT Press.

\bibitem[\protect\citename{Hinskens and Weijer}2003]{Hinskens:03}
F. Hinskens and J. Weijer.
\newblock 2003.
\newblock Patterns of segmental modification in consonant inventories: A cross-linguistic study. {\em Linguistics} 
{\bf 41}(6), 1041--1084.

\bibitem[\protect\citename{Jakobson}1968]{Jakob:68}
R. Jakobson.
\newblock 1968.
\newblock {\em Child Language, Aphasia and Phonological Universals}. The Hague: Mouton.

\bibitem[\protect\citename{Jeong \bgroup et al.\egroup}2000]{Jeong:00}
H. Jeong, B. Tombor, R. Albert, Z. N. Oltvai and A. L. Barab{\'a}si. 
\newblock 2000.
\newblock The large-scale organization of metabolic networks. {\em Nature} {\bf 406} 651–-654.

\bibitem[\protect\citename{Liljencrants and Lindblom}1972]{Lindblom:72}
J. Liljencrants and B. Lindblom.
\newblock 1972.
\newblock Numerical simulation of vowel quality systems: the role of perceptual contrast. {\em Language} {\bf 48}, 
839--862.

\bibitem[\protect\citename{Lindblom and Maddieson}1988]{Lindblom:88}
B. Lindblom and I. Maddieson. 
\newblock 1988.
\newblock Phonetic universals in consonant systems. {\em Language, Speech, and Mind}, 62--78, Routledge, London.

\bibitem[\protect\citename{Locke}1983]{Locke:83}
J. L. Locke.
\newblock 1983.
\newblock {\em Phonological Acquisition and Change}. Academic Press New York.

\bibitem[\protect\citename{Maddieson}1984]{Maddieson:84}
I. Maddieson.
\newblock  1984.
\newblock {\em Patterns of Sounds}, Cambridge University Press, Cambridge.

\bibitem[\protect{Mukherjee \bgroup et al.\egroup}{2007a}]{Mukherjee:06}
A. Mukherjee, M. Choudhury, A. Basu and N. Ganguly.
\newblock 2007a. 
\newblock Modeling the co-occurrence principles of the consonant inventories: A complex network approach. {\em Int. 
Jour. of Mod. Phys. C} {\bf 18}(2), 281--295.

\bibitem[\protect{Mukherjee \bgroup et al.\egroup}{2007b}]{acl:07}
A. Mukherjee, M. Choudhury, A. Basu and N. Ganguly.
\newblock 2007b.
\newblock Redundancy ratio: An invariant property of the consonant inventories of the world's languages {\em 
Proceedings of ACL07}, 104--111.

\bibitem[\protect\citename{Newman}2001]{Newman:01}
M. E. J. Newman.
\newblock 2001. 
\newblock Scientific collaboration networks. {\em Physical Review E} {\bf 64}, 016131.

\bibitem[\protect\citename{Newman}2003]{Newman:03}
M. E. J. Newman.
\newblock 2003.
\newblock The structure and function of complex networks. {\em SIAM Review} {\bf 45}, 167--256. 

\bibitem[\protect\citename{Peltom{\"a}ki and Alava}2006]{Alava:06}
M. Peltom{\"a}ki and M. Alava.
\newblock 2006.
\newblock Correlations in bipartite collaboration networks. {\em Journal of Statistical Mechanics: Theory and 
Experiment}, P01010.

\bibitem[\protect\citename{Ramasco \bgroup et al.\egroup}2004]{Ramasco:04}
J. J. Ramasco, S. N. Dorogovtsev and R. Pastor-Satorras.
\newblock 2004.
\newblock Self-organization of collaboration networks. {\em Physical Review E} {\bf 70}, 036106.

\bibitem[\protect\citename{Shannon and Weaver}1949]{Shan:49}
C. E. Shannon and W. Weaver.
\newblock 1949.
\newblock {\em The Mathematical Theory of Information}. University of Illinois Press, Urbana.

\bibitem[\protect\citename{Steels}2000]{Steels:00}
L. Steels. 
\newblock 2000.
\newblock Language as a complex adaptive system. In: Schoenauer, M., editor, {\em Proceedings of
PPSN VI}, {\em LNCS}, 17--26.

\bibitem[\protect\citename{Trubetzkoy}1931]{Trub:30}
N. Trubetzkoy. 
\newblock 1931.
\newblock Die phonologischen systeme. {\em TCLP} {\bf 4}, 96--116.

\bibitem[\protect\citename{Trubetzkoy}1939]{Trub:39}
N. Trubetzkoy.   
\newblock 1969.
\newblock {\em Principles of Phonology}. English translation of {\em Grundz{\"u}ge der Phonologie, 1939}, University 
of California Press, Berkeley.

\end{thebibliography}
\end{document}